\documentclass[apjl]{emulateapj}

\usepackage{xspace}

\def \ergsec{\ensuremath{\mathrm{erg}\,\mathrm{s}^{-1}}}

\def \chandra{\emph{Chandra}\xspace} 
\def \xmm{\emph{XMM-Newton}\xspace}

\def \ka{K$\alpha$\xspace}

\def \gcas{$\gamma$ Cas\xspace}

\shorttitle{\chandra observations of HD119682}
\shortauthors{Torrej\'on et al.}

\begin{document}


\title{Hot thermal X-ray emission from the Be star HD119682.}


\author{J.M. Torrej\'on\altaffilmark{1}, N.S. Schulz\altaffilmark{2}, M.A. Nowak\altaffilmark{2}, P. Testa\altaffilmark{3} and J.J. Rodes\altaffilmark{1} }

\affil{$^{1}$Instituto Universitario de F\'isica Aplicada a las Ciencias y las Tecnolog\'ias, Universidad de Alicante, E03080 Alicante, Spain; jmt@ua.es} 

\affil{$^{2}$MIT Kavli Institute for Astrophysics and Space Research, Cambridge MA 02139, USA}

\affil{$^{3}$Harvard Smithsonian Center for Astrophysics, Cambridge MA 02138, USA}




\begin{abstract}

We present an analysis of a series of four consecutive \chandra high resolution transmission gratings observations, amounting to a total of 150 ks, of the Be X-ray source HD119682 (= 1WGA J1346.5$-$6255), a member of the new class of $\gamma$ Cas analogs. The \chandra lightcurve shows significant brightness variations on timescales of hours. However, the spectral distribution appears rather stable within each observation and during the whole campaign. A detailed analysis is not able to detect any coherent pulsation up to a frequency of 0.05 Hz. The \chandra HETG spectrum seems to be devoid of any strong emission line, including Fe \ka fluorescence. The continuum is well described with the addition of two collisionally ionized plasmas of temperatures $kT\approx 15$ keV and 0.2 keV respectively, by the \texttt{apec} model. Models using photoionized plasma components (\texttt{mekal}) or non thermal components (\texttt{powerlaw}) give poorer fits, giving support to the pure thermal scenario. These two components are absorbed by a single column with $N_{\rm H}=(0.20^{+0.15}_{-0.03})\times 10^{22}$ cm$^{-2}$ compatible with the interstellar value. We conclude that HD119682 can be regarded as a \emph{pole-on} \gcas analog.

\end{abstract}


\section{Introduction} 

HD119682 (= 1WGA J1346.5$-$6255) is a Be star which presents copious X-ray emission. A member of the open cluster NGC 5281, located at a distance of $\sim 1.3$ kpc, this Be star displays an X-ray luminosity in the 0.5--10 keV energy band of the order of $L_{\rm X}\sim 10^{32}$ \ergsec, low when compared with the typical luminosities of Be X-ray binaries ($L_{\rm X}\sim 10^{36-37}$ \ergsec), two orders of magnitude lower than persistent Be X-ray binaries (Reig \& Roche, 1999) and at least one order of magnitude higher than the expected X-ray emission from isolated OB stars (Bergh\"{o}fer et al. 1997). With a mass of $M =18\pm 1M_{\odot}$ and an age of $4\pm1$ Myr, HD119682 seems to be a blue straggler (Marco et al. 2009), the second (out of the 7 known) gamma Cas analogs that shares this property. 

Rakowski et al. (2006; R06), using the superb spatial resolution of \chandra\ as well as H$\alpha$ images from the Clay 6.5 m Telescope in the Magellan Observatory at Las Campanas, unambiguously identified the X-ray source 1WGA J1346.5$-$6255 with the OB star HD119682. Using \xmm\ {\it PN+MOS} data, these authors showed that the X-ray spectrum is well described by the addition of two optically thin (\texttt{raymond}) thermal plasmas with $kT_{1}=1.07\pm 0.16$ keV and $kT_{2}=10.4^{+2.3}_{-1.6}$ keV respectively. Based on the similarities shared between HD119682 and \gcas\ (early type optical counterpart, thermal nature of the X-ray emission, $\log L_{X}/L_{\rm bol}=-5.4$, etc) R06 classify the source as a \gcas\ analog, and argue against other X-ray production mechanisms like colliding winds (with an unseen companion) or emission from a magnetically confined wind (given the very high temperature of the hot plasma).

Safi-Harb et al. (2007; S07), in a multifrequecy analysis, confirm the identification between 1WGA J1346.5$-$6255 and HD119682 and characterize the optical counterpart (B0.5Ve), definitely unrelating this source from the background Supernova Remnant G309.2-00.6.  These authors find that the \xmm\ {\it PN+MOS} data can be satisfactorily described by the addition of two optically thin thermal plasmas, using the \texttt{mekal} model,  with $kT_{1}=1.7\pm 0.3$ keV and $kT_{2}=13.0^{+2.6}_{-2.4}$ keV respectively, while the fit to the \chandra\ spectrum, extracted from the S3 ACIS CCD chip, delivers temperatures of $kT_{1}=0.95^{+0.34}_{-0.24}$ keV and $kT_{2}\geq 14$ keV respectively. Alternatively, the hotter thermal component can be described by a powerlaw. This degeneracy has also been found in \gcas (Smith et al. 2004, Lopes de Oliveira et al. 2011) and HD110432 (Lopes de Oliveira 2007, Torrej\'on et al. 2012). S07 present also hints of a possible pulsation with $P\simeq 1500$ s which could be caused by the spin of a NS or a WD.

The origin of the X-ray emission in \gcas analogs is still a mystery. Two hypotheses have been put forward:
\begin{itemize}
\item The X-ray binary scenario in which a White Dwarf (WD) accretes matter from the Be donor (Haberl, 1995; Kubo et al., 1998; Owens et al., 1999 for the case of \gcas; Torrej\'on \& Orr, 2001 for HD110432). The low potential well of the WD, compared with that of a NS, accounts naturally for the lower luminosity and the thermal nature of the X-ray emission. This explanation faces several problems. On one hand, the X-ray emission shows no sign of any orbital modulation or spin pulse. On the other hand, the progenitor of the putative WD should have been still more massive than the early B companion ($M\sim 20 M_{\odot}$). From the evolutionary perspective, it is challenging to produce a WD from such a massive star; naturally, they would explode as a supernova leaving behind a NS, as is the case for the Be X-ray binaries. In the case of HD119682 no SN explosion seems to have occurred (Marco et al. 2009) which argues against the NS companion.

\item The standalone star scenario in which the X-rays are produced by the interaction of the star's magnetic field with the circumstellar disk (Robinson et al. 2002, Smith \& Robinson 2003, Smith et al. 2004). In this scenario, magneto rotational instabilities in the inner part of the circumstellar disk around the Be star (close to the Keplerian co-rotation radius) enhance a seed magnetic field through a disk dynamo mechanism (Balbus \& Hawley, 1998). The differential rotation of the disk, stretch and sever the entangled field lines accelerating particles which, upon impact on the star's surface, heat the plasma up to MK temperatures. The finding that the X-ray emission seems to be produced close to the surface of the Be star supports this scenario (Smith et al. 2012, Torrej\'on et al. 2012). The details of this mechanism, though, are far from clear. Why other early type Be stars with large circumstellar disks do not produce X-ray emission at all is not explained. 
\end{itemize}

In any case, to understand the true nature of these systems will have important consequences in our picture of the structure and evolution of early type stars and/or how binary systems evolve.

In this work we present an analysis of a series of four consecutive \chandra High Energy Transmission Gratings (HETG) observations of the Be star HD119682 allowing us to study the spectrum at the highest resolution available today. The four observations amount to a total of $\sim 150$ ks, spread over five days, allowing a sensitive search for pulsations in a wide range of frequencies.

\section{Observations}

\begin{deluxetable}{rccc}
\tabletypesize{\scriptsize}
\tablecaption{Journal of \chandra HETG observations.} 
\tablewidth{0pt}
\tablehead{
 \colhead{ObsID} & \colhead{Date} & \colhead{$t_{\rm exp}$} & \colhead{Rate$^{a}$}\\
& (dd-mm-yyyy) & (ks) & (c/s) \\
}
\startdata
 8929 & 17-12-2008 00:20:10 & 28.9  &  0.039  \\
 10835 & 19-12-2008 09:32:42  & 29.7   & 0.031 \\
 10834& 20-12-2008 15:45:04 & 59.3 & 0.032 \\
 10836 & 21-12-2008 15:47:03 & 29.9 & 0.035
\enddata
\tablenotetext{a}{Zeroth order}
\label{tab:obs}
\end{deluxetable}


We have processed and analyzed the \chandra data from the ObsIDs 8929, 10835, 10834 and 10836 available at the public archive (PI C.E. Rakowski). In Table \ref{tab:obs} we show the journal of observations. The \chandra High Energy Transmission Gratings spectrometer (HETG; Canizares et al. 2005) observed the source  during 150 ks, splitted in four intervals. Both HETG instruments, the High Energy Gratings (HEG; 0.7-8 keV) and Medium Energy Gratings (MEG; 0.4-8 keV) were used for the analysis. After filtering the data, the 0.5 to 7.5 keV spectral range showed enough signal-to-noise ratio for the scientific analysis. Both the spectra as well as the response files (arf and rmf)
were extracted using the CIAO software (v 4.4). First dispersion orders ($m=\pm 1$) were fitted
simultaneously. The extracted spectra were binned to
match the resolution of HEG (0.012 \AA\ {\it FWHM}) and MEG (0.023
\AA\ {\it FWHM}) and have a minimum S/N ratio of 7. The peak count rate in a single (MEG) gratings arm spectrum is $\approx $ 0.008 cts/sec/\AA\ and, consequently, the spectra is not affected by pileup\footnote{See \emph{The Chandra ABC Guide to Pileup}, v.2.2, \texttt{http://cxc.harvard.edu/ciao/download/doc/pileup-abc.pdf}}. For the lightcurve extractions we add data from all dispersed orders ($m=-3, -2, -1, 1, 2, 3$) from HEG and MEG. High dispersion orders ($m=\pm 3, \pm 2$) constitute a 25 \% of the total count rate.

The analysis was performed with the 
Interactive Spectral Interpretation System (ISIS) v 1.6.1-24 \citep{houck00}. 
The photoelectric absorption has been treated with the \texttt{tbnew} absorption model (Wilms et al. 2012), which includes the most up to date photoelectric cross sections.   


\begin{figure*}
\includegraphics[angle=0,width=\columnwidth]{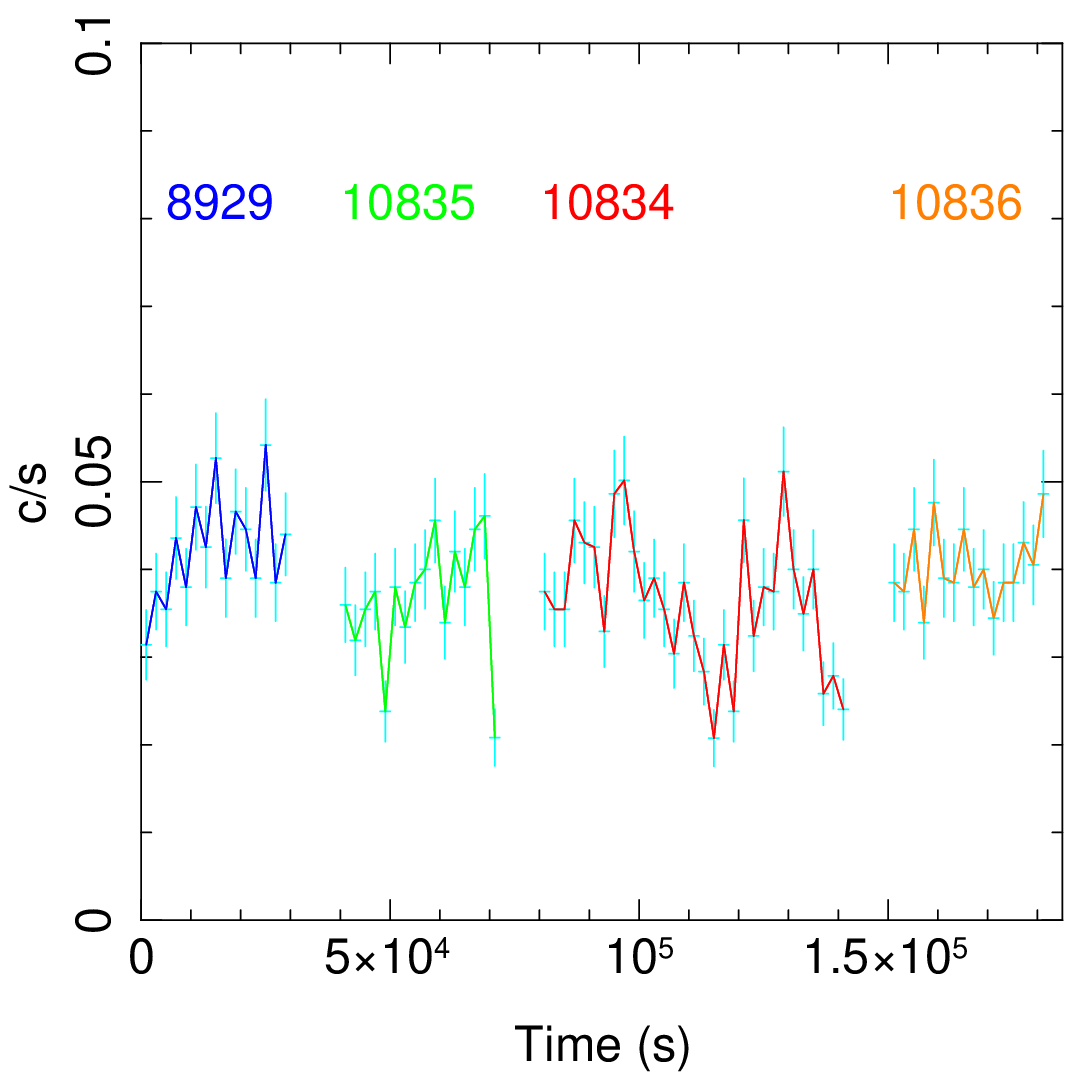}
\includegraphics[angle=0,width=\columnwidth]{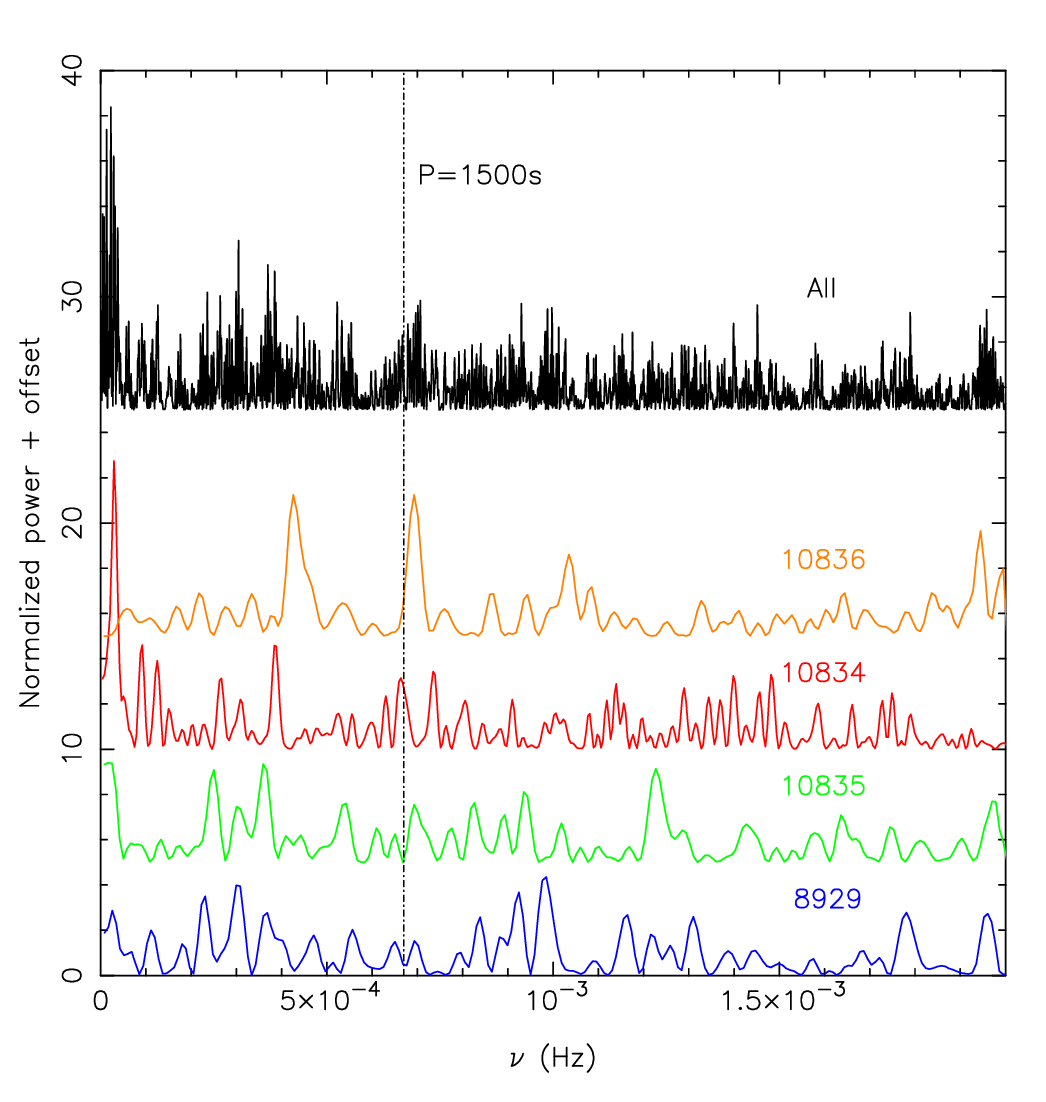}
\caption{Left: \chandra lightcurve, in the 2 - 25 \AA\ wavelength range, binned in 2000 s intervals. All lightcurves have an arbitrary time offset and, therefore, the time gaps are not to scale. Right: Power spectrum corresponding to the 250 s binned \chandra HETG lightcurve. The 1500 s pulsation ($\nu\simeq 6.7\times 10^{-4}$ Hz) is only detected in ObsID 10836 but is not detected neither in the rest of the ObsIDs nor in the combined ligthcurve. }
\label{fig:chandra_lc}
\end{figure*}

\section{Timing analysis}

The four observations analyzed in this work span five different days with uninterrupted coverages ranging from 30 ks to 60 ks. In total they represent 150 ks worth of data. This allows us to search for coherent pulsations over a wide range of frequencies with unprecedented detail. In the left panel of Fig. \ref{fig:chandra_lc} we present the \chandra lightcurve binned to 2000 s bins. As can be seen, the source brightness is variable on timescales shorter than the duration of a single observation (ObsID). During the whole campaign, however, the source remains quite stable. In order to search for coherent pulsations we have analyzed the lightcurve with a maximum time resolution of 10 s. Subsequently we have created lightcurves with larger time bins of 100, 250 and 500 s to reduce the error bars. In Fig. \ref{fig:chandra_lc}, right panel, we show, as an example, the normalized Fast Fourier Transform periodogram of the lightcurve using 250 s bins, using the Lomb - Scargle technique, for each ObsID. The upper plot corresponds to all four intervals together taking into account the duration of the time gaps between each observation. Safi-Harb et al. (2007) suggested the possibility of a $\sim 1500$ s pulsation in the lightcurve of HD119682. In our last ObsID (10386) there is indeed a peak at $P=1443\pm 17$ s (and also other at $2350\pm 50$ s). The significance of this peak, as given by the False Alarm Probability (FAP), is low. FAP is defined as the probability that no periodic component is present in the data with such a period within the current frequency search parameters. In order to compute this significance we have run $10^{3}$ Monte Carlo simulations of the randomized Y-axis values of the lightcurve and performed the corresponding periodograms. Then we search for peaks higher than that in the original periodogram at any frequency. We have found higher peaks in $20\pm 3$ \% (two sigma standard error) of the periodograms. Therefore, the False Alarm Probability is rather high, $\sim 20\%$ and, consequently, the significance of the peak low. Similar results are obtained irrespectively of the size of the time bin. However neither of these peaks are present in the previous data sets. Although some low significance peaks can be seen in each periodogram, none of them remain stable. In consequence, in the whole data set we do not find evidence of any significant period up to a frequency of 0.05 mHz\footnote{The Nyquist frequency corresponding to the 10 s time bin.}).

\begin{figure}
\includegraphics[angle=0,width=\columnwidth]{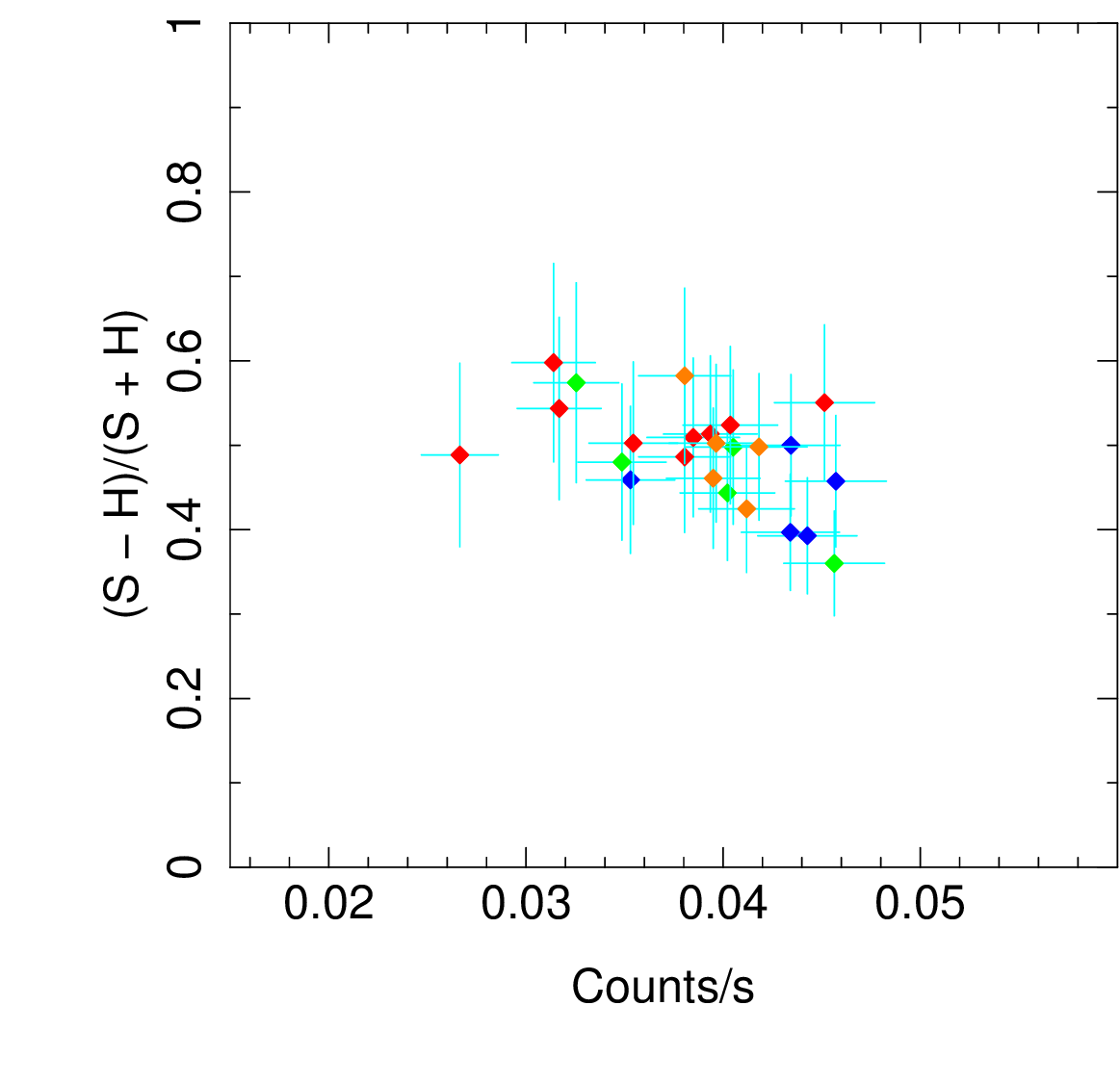}
\caption{\chandra hardness-intensity plot using 7000 s time bins. Different colors represent the different ObsIDs, with the same color coding as in Fig. \ref{fig:chandra_lc}. While changes in brightness from bin to bin are, in general, significant the hardness ratios are consistent with a constant value.}
\label{fig:colors}
\end{figure}

In order to study the spectral variations of the X-ray source we have extracted the \chandra lightcurves in the wavelength intervals 3--6 keV (H), 2--3 keV (M) and 0.5--2 keV (S), and have created the corresponding hardness ratios versus source intensity plots. In Fig. \ref{fig:colors}, we show the variation of the soft/high hardness ratio against the source intensity (0.5--6 keV) using 7000 s time bins. As can be seen, the source does not show any spectral correlation with brightness. The average values of the hardness ratios are $0.43\pm 0.07$ (ObsID 8929), $0.48\pm 0.07$ (ObsID 10835), $0.53\pm 0.06$ (ObsID 10834) and $0.50\pm 0.06$ (ObsID 10836). They are, therefore, compatible within errors. The differences between single values of the hardness ratio are not significant either. Individual errors range between 0.15 and 0.20. In conclusion, while changes in brightness from bin to bin are, overall, significant the hardness ratios are consistent with a constant value. The source appears rather monochromatic on timescales from hours to days.

\section{Spectral analysis}


\begin{deluxetable}{llcc}
\tabletypesize{\scriptsize}
\tablecaption{Model parameters for \chandra\ HETG  data. } 
\tablewidth{0pt}
\tablehead{
  \colhead{Component}  & \colhead{Parameter} &  & \colhead{Value} 
}
\startdata
     &        & PL+APEC & 2 APEC \\
     &        &    &   \\
     & $N^{a}_{\rm H}$  & 0.15$^{+0.15}_{-0.03}$ & 0.20$^{+0.15}_{-0.03}$ \\
     &        &    &   \\
powerlaw         & Norm & 0.00026$^{+0.0005}_{-0.0004}$ & ...\\
                 & $\Gamma$ & 1.36$^{+0.15}_{-0.01}$ & ...\\
                 &          &     &  \\
apec 1            & Norm &  ... & 0.0018$^{+0.0005}_{-0.0006}$ \\
            & $kT$ (keV) & ... & 15.7$^{+4.7}_{-5.4}$ \\
            &        &      &  \\
apec 2      & Norm & 0.0003$^{+0.0003}_{-0.0001}$ & 0.0005$^{+0.0001}_{-0.0002}$ \\
            & $kT$ (keV) & 0.21$^{+0.02}_{-0.03}$ & 0.20$^{+0.03}_{-0.02}$ \\
            &         &           &   \\ 
            & Flux$^{b}$   &  1.78    & 1.81  \\
            &        &    &   \\
            &  $\chi^{2}_{\rm r}$ (dof)  & 1.48 (86)   & 1.35 (86)    
\enddata
\tablenotetext{a}{In units of $\times 10^{22}$ cm$^{-2}$}\\
\tablenotetext{b}{Unabsorbed 0.5--8 keV flux in units of $\times 10^{-12}$ erg s$^{-1}$ cm$^{-2}$}
\label{tab:results}
\end{deluxetable}

\begin{figure}
\includegraphics[angle=-90,width=\columnwidth]{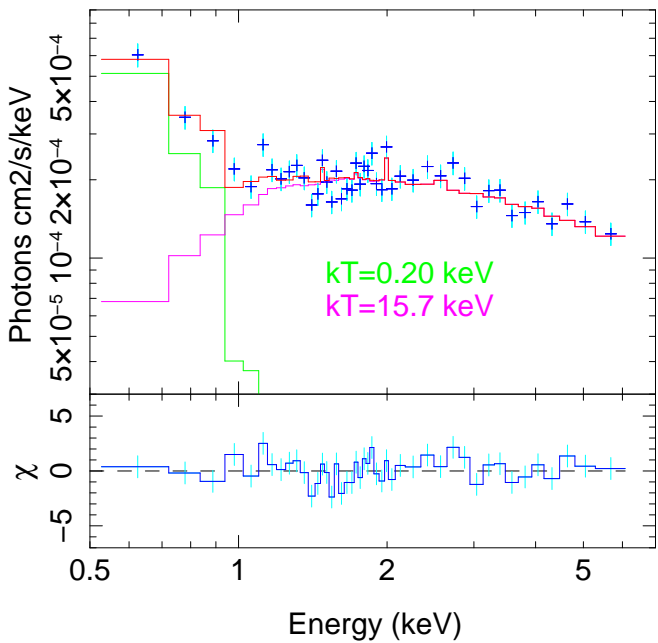}
\caption{The \chandra {\it HETG} unfolded spectrum of HD119682 showing the relative contributions of the low and high temperature plasmas respectively. Also shown (in red) is the best fit model.}
\label{fig:nolines}
\end{figure}

According to the previous analysis, the spectra of HD119682 is fairly stable in a time scale of days despite the variations in the brightness. Therefore we have combined the four observations into a single spectrum to increase the signal to noise ratio ($S/N$). The resulting spectrum is shown in Fig. \ref{fig:nolines}.

In order to describe the continuum, models consisting of one component do not give satisfactory results. For example, an absorbed \texttt{powerlaw} gives a reduced chi square $\chi^{2}_{r}=1.71$ for 88 degrees of freedom (dof) while a single \texttt{mekal} gives  $\chi^{2}_{r}=1.64$ (88 dof). The spectrum shows always a pronounced excess below 1 keV which can only be modeled when a second component to the model is added. The best fit is attained with the addition of two optically thin collisionally ionized plasmas (\texttt{apec}). The parameters are shown in the fourth column of Table \ref{tab:results}. The low energy range is fitted by a plasma with $kT = 0.20^{+0.03}_{-0.02}$ keV and solar abundance. The high sensitivity of the \chandra gratings allow us to constrain this temperature with unprecedented accuracy. Although there are few energy bins below 1 keV, the associated errors (blue crosses) are very small. The obtained plasma temperature $kT_{2}= 0.20^{+0.03}_{-0.02}$ keV is fully compatible with the temperature of the cool plasma in HD110432 derived from high resolution spectra (Torrej\'on et al. 2012) and significantly lower than that obtained by R06 ($kT= 1.07\pm 0.16$ keV; Raymond-Smith) and S07 ($kT= 1.7\pm 0.3$ keV; mekal). The absorption column, however, remains essentially compatible (see below). Therefore, the differences in $kT$ could be due to intrinsic long term variability of the source. This variability in the cool plasma temperature seems also to be present in \gcas (Lopes de Oliveira et al. 2010) while the hot plasma temperature appears to be more stable.

The high energy range requires a second, much hotter \texttt{apec} model with $kT=15.7^{+4.7}_{-5.4}$ keV. This temperature is compatible within the errors with that derived by R06 and S07. Alternatively, in agreement with previous works on \gcas candidates, the high energy range can be fit with a \texttt{powerlaw} with a photon index $\Gamma=1.36$. The parameters are shown in the third column of Table \ref{tab:results}. However, in terms of $\chi^{2}$ the two \texttt{apec} model is significantly better.  Replacing the collisionally ionized plasmas (\texttt{apec}) by photoionized plasmas (\texttt{mekal}), gives slightly worse fits. For example, the two \texttt{mekal} model gives $\chi^{2}_{r}=1.38$ (86 dof) while \texttt{powerlaw + mekal} gives $\chi^{2}_{r}=1.49$ (86 dof). In conclusion, the fully thermal model composed by two \texttt{apec} components is favored by the present data set. 

The most conspicuous characteristic of the spectrum is the lack of any discernible emission line. In particular, no Fe \ka line is seen and even the flux at 6 keV is below the 7 $\sigma$ level. There is flux up to 7 keV at the 3$\sigma$ level; however, there is certainly not an Fe emission line even at 1 $\sigma$. Finally, the residuals present some structure between 1 and 2 keV that could be due to the presence of emission lines. Indeed, this is the line rich region for highly ionized plasmas. The HETG spectra of bright \gcas analogs display several strong emission lines in this region (Smith et al. 2004, Fig. 2; , for \gcas; Torrej\'on et al. 2012, Fig. 7 and 8, for HD110432). Strong lines of Ne \textsc{x} Ly$\alpha$ (1.02 keV) and Si \textsc{xiv} Ly$\alpha$ (2.00 keV) as well as weaker lines of Fe \textsc{xxiv} (1.17 keV) and Mg \textsc{xii} Ly$\alpha$ (1.47 keV) might be expected in this area. However, none of these lines show up in the spectrum of HD119682 with EW upper limits of the order of 1 eV.

The high sensitivity of \chandra HETG at low energies also allows us to study the absorption column with unprecedented detail. The analysis of brighter members of this class have shown that the different spectral components are affected by different absorption columns (Smith et al. 2004, for \gcas; LO07 and Torrej\'on et al. 2012, for HD110432). Here we have tried the same strategy. The values of the resulting columns turn out to be identical, within the errors, for all models. Therefore a single photoelectric absorption satisfactorily describes the spectrum of HD119682. The measured $N_{H}=(0.20^{+0.15}_{-0.03})\times 10^{22}$ cm$^{-2}$ is very low, in agreement with R06 and S07.

\section{Discussion}


The spectrum of HD119682 differs from other \gcas analogs observed at high resolution mainly in that it shows no discernible emission lines. Particularly striking is the fact that no Fe complex is observed. Despite this, the continuum spectral parameters of HD119682 are very similar to those of \gcas (Smith et al. 2004, Fig. 2) and HD110432 (Torrej\'on et al. 2012). Consistently with these previous works, the correct description of the spectrum is attained with the addition of several thermal components. In the case of HD119682 only two are needed with characteristic temperatures in the $\sim 2$ MK and $\sim 150$ MK range. Both components are modified by a single absorption column $N^{X}_{\rm H}\sim 0.20\times 10^{22}$ cm$^{-2}$. This column would imply an interstellar excess $E(B-V)=N^{X}_{\rm H}/6.83\times 10^{21}=0.29^{+0.03}_{-0.01}$ (Ryter 1996\footnote{The errors represent the propagated uncertainty in the measured $N^{X}_{\rm H}$}) in good agreement with the interstellar medium (ISM) absorption towards NGC 5281 (Moffat \& Vogt 1973, $E(B-V)=0.26\pm 0.03$; Marco et al. 2009, $E(B-V)=0.28\pm 0.02$).

HD119682 has been observed several times both in X-rays (\xmm\ August 2001 and \chandra\ December 2004, R06, S07; December 2008, this work)  and in the H$\alpha$ light (June 2003 and February 2004; R06, S07) showing always a strong emission. Although the time coverage is still poor, the source appears to be persistent in both energy bands. Assuming that the circumstellar envelope (and hence the H$\alpha$ emission) was also present at the epoch of our \chandra observation, the X-ray source was not obscured by the disk material. The same would be true for the observations analyzed in previous works (R06, S07). The circumstellar disk (confined to the equatorial plane) does not intercept the X-ray emission. In the framework of the accreting binary system, the compact object (the WD) should not be deeply embedded into the circumstellar disk; yet it should be accreting enough material to power persistently the X-ray source up to a non negligible $L_{X}\sim 3.8\times 10^{32}$ \ergsec. This would require a very special tuning of the astrophysical parameters of the system. Yet, even though rare, this situation can happen: 4U2206$+$54 is a low luminosity ($L_{\rm X}\approx 10^{35-36}$ \ergsec) HMXB with a NS accreting from the wind of a massive O9.5V star while the X-ray spectrum presents very few (if any) local absorption (Reig et al. 2012). Contrary to the case of 4U2206$+$54, however, no pulsations have been found for HD119682 or any other \gcas analog so far. In contrast, within the framework of the standalone active star, the single absorption column, compatible with the ISM value, can be simply understood if the system is seen at very high inclination angles. It has been recently found that the hard X-ray component of the multi-thermal emission of \gcas (Smith et al. 2012) and HD110432 (Torrej\'on et al. 2012) is very likely produced near the surface of the Be star and is absorbed by material with a column density compatible with the circumstelar disk. If this is also correct for HD119682, then the $N_{H}^{X}\approx N_{H}^{ISM}$ implies that the system must be observed at high inclination. This would be consistent with the single peaked emission lines seen in the optical spectrum of this object (S07).


R06 and S07 report hints of Fe \ka emission. However, no sign of such emission line (nor any other) is present here. Thus, the fluorescence emission from neutral Fe seems to be variable in the long term. On the other hand the 0.5-8.5 keV X-ray luminosity of the source has varied within a maximum factor of $\sim 2$ in a period of seven years (R06, S07, this work). One of the possible sites for the \ka\ reprocessing of the X-rays is the circumstelar disk. Smith et al. (2012) find a direct correlation between the absorption column of the hard emission and the EW(Fe \ka) (as well as with the reddening) for \gcas (Table 5 of that reference) implying that most, perhaps all, of the X-ray emission is produced {\it behind} the circumstellar disk material. These authors also deduced an inclination of $i=[43\pm 1]^{\rm o}$ for \gcas. In the case of HD119682, however, the absorption column remains always compatible with the ISM value despite the changes in the X-ray brightness of the source and variable Fe \ka emission because the circumstellar disk, confined to the equatorial plane, never intercepts the X-ray emission. 
 
HD119682 would thus appear as a {\it pole-on} \gcas, an important benchmark to explore the phenomenon where the plasma, and its variations, could be directly observed without any intervening circumstellar material.

\acknowledgments

The authors acknowledge the constructive criticism of an anonymous referee. This work has been supported by the Spanish
Ministerio de Ciencia e Innovaci\'on (MICINN) through the grants AYA2010-15431 and AIB2010DE-00057.  


\end{document}